\documentclass[twocolumn,secnumarabic,amssymb, nobibnotes, aps, prl]{revtex4-1}
\usepackage{comment}
\usepackage[dvipdfmx]{graphicx}
\begin{document}
\title{Group Formation through Indirect Reciprocity}
\author{Koji Oishi}
\author{Takashi Shimada}
\author{Nobuyasu Ito}
\affiliation{Department of Applied Physics, School of Engineering, The University of Tokyo, 7-3-1, Hongo, Bunkyo-ku, Tokyo, 113-8656, Japan.}
\pacs{89.75.Fb}
\begin{abstract}
The emergence of structure in cooperative relation is studied in a game theoretical model.
It is proved that specific types of reciprocity norm lead individuals to split into two groups.
The condition for the evolutionary stability of the norms is also revealed.
This result suggests a connection between group formation and a specific type of reciprocity norm in our society.
\end{abstract}
\maketitle
Cooperative behavior can be observed widely in our society.
An important aspect of cooperation is that people can maintain mutual cooperation even if they are faced with social dilemma, which means the situation that people fail to cooperate although entire cooperation gives better payoff to everyone.
Several mechanisms are pointed out to be important to overcome the social dilemma, and reciprocity is one of them~\cite{Nowak}.
Especially indirect reciprocity (cooperation to someone enhance his probability to have a cooperative response from others) is a key when partners of interaction are flexible~\cite{Nowak-Sigmund-1998}.
It is necessary for this mechanism to work that people monitor what others do to others and appropriately judge whom they should cooperate with.
Otherwise free riding behavior spreads and cooperation is abandoned.

Another important aspect of cooperation is that people are generally not equally cooperative to all people.
Especially, they often form several groups and cooperate only inside of the groups, which is called in-group favoritism~\cite{Tajfel}.
To understand in-group favoritism, one can assume a given group structure and study how cooperation is sustained only inside of the groups~\cite{Masuda-Ohtsuki,Masuda,Nakamura-Masuda}.
In some cases, however, the group structure itself can be formed by social interaction of people.
For example, groups of friends or business partners are hardly decided \emph{a priori}.
Therefore a model to explain in-group favoritism without exogenous group structure is required.

Here, we study whether group structure of cooperation is spontaneously formed in a model of indirect reciprocity with private assessment: each person can have different opinions on the others' reputation.
Private assessment model itself has been studied in previous works~\cite{Brandt-Sigmund-2004,Takahashi-Mashima,Uchida,Sigmund-2011}.
However their main attention was not the emergence of group structure but what definition of goodness can evolutionary promote cooperation.
In this paper, on the other hand, the structure of the cooperative relation of the main interest.

We consider well-mixed $N$ players playing a game, which is called donation game.
The number of players is assumed to be large, $N \gg 1$.
The game is repeated over a large number of rounds.
In every round, one player is randomly chosen as a \emph{donor} and another player is randomly chosen as a \emph{recipient}.
The donor decides whether he cooperates or defects.
If the donor cooperates, the payoff of the donor in the round is $-c<0$ and that of the recipient is $b>0$.
If the donor defects, payoffs of both players are 0.
$b$ and $c$ are called benefit of cooperation and cost of cooperation respectively.
We assume $b>c$, which means players are faced with social dilemma.

Each player has his own opinion on each of all players whether he is good or bad.
The donor of each round decides his action based on his opinion on the recipient of the round.
All players observe who is the donor, who is the recipient, and what the donor did to the recipient.
Then all players independently revise their own opinion on the donor based on the observation in the round.
A round ends after all players revise their opinion.

$\beta_{ij}(t) \in \{G,B\}$ represents player $i$'s opinion on player $j$ at the outset of round $t$.
If player $i$ has the opinion that player $j$ is good at the outset of round $t$ then $\beta_{ij}(t)=G$ and if $i$ has the opinion that $j$ is bad then $\beta_{ij}(t)=B$.
In this paper, the matrix $\beta(t)=\{ \beta_{ij}(t) \}_{i,j}$ is called opinion matrix, which was called image matrix in previous works~\cite{Uchida,Sigmund-2011}.
Each element of the opinion matrix at initial round $t=1$, is $G$ with probability $p$ and $B$ with probability $1-p$.
The probability $p$ is called initial trust probability and assumed it is not $0$ or $1$.
Note that each player $i$ also has opinion of himself $\beta_{ii}$, which can matter when he is the recipient and revises his opinion on the donor.

The action matrix $\gamma(t)=\{\gamma_{ij}(t)\}_{i \neq j}$ represents the action each player will take to each of other players if they are matched at round $t$.
If $i$ will cooperate with $j$, $\gamma_{ij}(t)=C$ and we say $i$ is cooperative to $j$, while if $i$ will defect against $j$, $\gamma_{ij}(t)=D$ and we say $i$ is not cooperative to $j$.
Diagonal elements are not defined since no one plays the game with himself.
We omit the the argument $(t)$ when it is not confusing.

The deterministic map from one's opinion on the opponent to the action he will take is called ``action matrix''.
Since the number of the possible opinions and that of possible actions are two, the number of the possible action rules are $4=2^2$.
The four action rules are called DISC, pDISC, All-C, and All-D, and those who adopt each action rule are called DISC players, pDISC players, All-C players, and All-D players, respectively.
All-C players always cooperate and All-D players always do not cooperate regardless of their opinion on the recipient.
In contrast, DISC players cooperate only when they regard the recipient as good while pDISC players cooperate only with bad recipients in their opinion.

Players also adopt certain ``assessment rule'', which define their opinions on the donor after observing a game round.
There are $2 \times 2$ possible combinations of an observer's opinion on the recipient and the action of the donor.
Therefore, there are $2^4$ assessment rules, which are denoted type 1 to type 16.
The left part of TABLE \ref{assessment_rules} shows the definition of the sixteen assessment rules.
Players do not change their action rule and assessment rule in repetition of games.
In this paper, all players are assumed to adopt a common assessment rule.
\begin{table}
    \caption{
      Definition of 16 assessment rules (2nd to 5th columns) and corresponding stationary states of the action matrix with each assessment rule and DISC or pDISC action rules (6th and 7th columns). 
      Each element in the 2nd to 5th columns shows the observer's new opinion based on his observation: (recipient's reputation (G/B), donor's action (C/D)).
      The classification of the state form (a) to (h) is described in the FIG. \ref{stationary_states}.
    }
    \label{assessment_rules}
    \begin{ruledtabular}
        \begin{tabular}{c|cccc|cc}
            rule&$(G,C)$&$(G,D)$&$(B,C)$&$(B,D)$&DISC&pDISC\\
            1&$B$&$B$&$B$&$B$&(b)&(a)\\
            2&$B$&$B$&$B$&$G$&(e)&(a)\\
            3&$B$&$B$&$G$&$B$&(b)&(e)\\
            4&$B$&$B$&$G$&$G$&(h)&(h)\\
            5&$B$&$G$&$B$&$B$&(b)&(a)\\
            6&$B$&$G$&$B$&$G$&(e)&(a)/(b)\\
            7&$B$&$G$&$G$&$B$&(g)&(g)\\
            8&$B$&$G$&$G$&$G$&(e)&(b)\\
            9&$G$&$B$&$B$&$B$&(b)&(a)\\
            10&$G$&$B$&$B$&$G$&(f)&(f)\\
            11&$G$&$B$&$G$&$B$&(a)/(b)&(e)\\
            12&$G$&$B$&$G$&$G$&(a)&(e)\\
            13&$G$&$G$&$B$&$B$&(c)&(d)\\
            14&$G$&$G$&$B$&$G$&(a)&(b)\\
            15&$G$&$G$&$G$&$B$&(a)&(b)\\
            16&$G$&$G$&$G$&$G$&(a)&(b)\\
        \end{tabular}
    \end{ruledtabular}
\end{table}

\begin{figure}[!htb]
    \includegraphics[width=8cm]{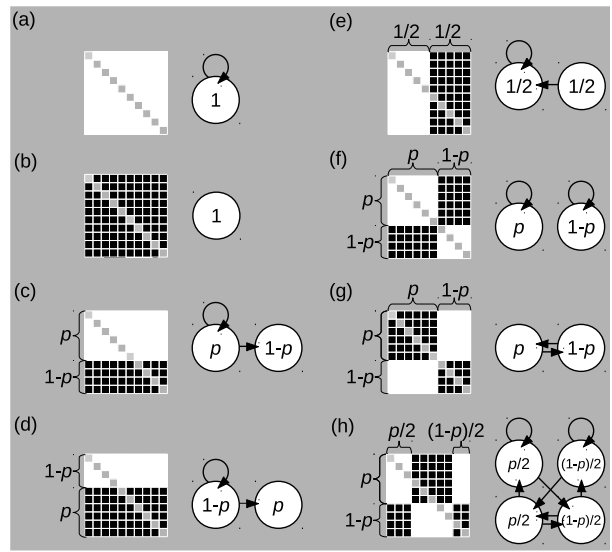}
    \caption{
      Stationary states are shown, by the corresponding action matrices (left in the each panel) and the schematic diagram of clustering structure (right).
      Elements of action matrices are shown with white (cooperative) and black (not cooperative) pixels.
      Diagonal pixels are grey since they are not defined.
      Matrices shows are sorted to make their structure easier to see. 
      Circles in schematic diagrams represents clusters of players.
      Arrows show players in which cluster are cooperative to players in which cluster.
      The fraction of the size of each cluster over total population is shown in each circle.
      The $p$ which appears in the fractions is the initial trust probability.
    }
    \label{stationary_states}
\end{figure}
First, we assume all players adopt a common action rule.
Under this restriction, we still have $16 \times 4$ cases for combinations of an assessment rule and an action rule.
Although the time development of an action matrices is stochastic and complicated, stationary states exist and they can be derived rigidly. 
Then it is proved that all stationary states can be classified into eight distinct states, according to the clustering structure of the action matrix.
FIG. \ref{stationary_states} shows the eight stationary states, which are denoted state (a) to (h).
The sizes of the clusters can be stochastic variable in some states, however we neglect the fluctuation since $N$ is assumed to be very large.
If all players in a cluster are cooperative only within that cluster, we call the cluster an exclusive group.
Among the eight states, only state (f) has multiple exclusive groups.

When the action rule is All-C, the stationary state is (a) irrespective of the assessment rule: all the players always cooperate with all the others in the end.
And players always only defect each other in the stationary state (that corresponds to state (b)) if they adopt All-D for the action rule.
TABLE \ref{assessment_rules} shows the stationary states in the case the action rule is DISC or pDISC.
It is worth stressing that the most interesting stationary state (f), which has two exclusive groups in it, appears only when players adopt type 10 assessment rule.

The type 10 assessment rule, which is called Kandori~\cite{Kandori} and hereafter denoted as KN, is well studied.
It is well known that KN promotes cooperation if players always share the same opinion, which is called public assessment~\cite{Kandori,Ohtsuki-Iwasa-2004,Ohtsuki-Iwasa-2006-a,Ohtsuki-Iwasa-2006-b}
It is also known that cooperation is less promoted by KN in private assessment ~\cite{Uchida,Sigmund-2011}.
However, it is first pointed out here that KN (and only KN) cause spontaneously split of players into multiple exclusive groups.

The essence of the derivation of the stationary states is as the following.
Let us focus on the case that the action rule is DISC or pDISC, since it is trivial that the action rule All-C yields state (a) and All-D gives state (b).
We first discuss the dynamics of opinion matrices and derive their stationary structure.

At initial state, the opinions of any two players have no correlation, therefore disagreements of opinions on some players almost always exist between any two players.
However, as we will see, agreements among at least part of players are formed in the repetition of the game.

\begin{figure}[!htb]
    \includegraphics[width=9cm]{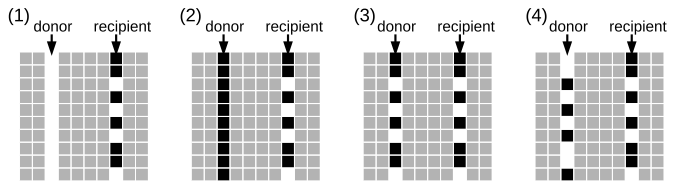}
    \caption{
      Four possible opinion matrices \emph{after} a revision.
      Elements of opinion matrices are shown with white (good) and black (bad) pixels.
      Elements irrelevant with the donor and the recipient do not affect, therefore they are shown with grey pixels.
      (1), (2), (3), and (4) correspond to the situation that all the players regard the donor as good, the all regard the donor as good, the all have the same opinion on the donor and on the recipient, and the all have the opposite opinions on the donor and on the recipient, respectively.
      Note that diagonal elements are also defined since they are not action matrices but \emph{opinion} matrices.
    }
    \label{matrix}
\end{figure}

Let us consider how (partial) agreement is formed during a round. 
Each player observing a round  revises his opinion on the donor according to the assessment rule, which is the function of his opinion on the recipient and the action of the donor.
Since the action of the donor is commonly observed, only the difference in the opinion on the recipient can cause the difference in the new opinion on the donor.
And some assessment rules, given an action of the donor, return the same reputation about the donor irrespective to the recipient's reputation.
Therefore, the new opinions of observers on the donor (column in the opinion matrix in FIG \ref{matrix}) after the round is either uniformly good/bad or copy/inversion image of the observers' opinion on the recipient.

\begin{figure}[!htb] 
    \includegraphics[width=5cm]{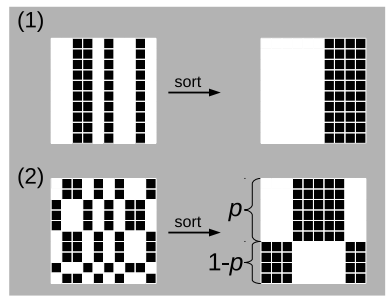}
    \caption{
      Opinion matrices of players who form (1) one opinion group or (2) two opinion groups with size ratio $p:1-p$.
      Opinion matrices are shown with white (good) and black (bad) pixels.
      Matrices in right hand side are results of sorted to make their structure easier to see.
    }
    \label{final_opinion}
\end{figure}

Under 12 out of 16 assessment rules, at least one of the actions of the donor (cooperation/defection) determines the new opinion of the observers irrespective to their opinion on the recipient.
It means that the number of players on whom all players share the same opinion can increase.
Once all players share the same opinion on all players, they never disagree about any players after the round.
Therefore, players finally reach and are fixed to the state that all players share the same opinion on all players.
In such cases, we say all players are in one opinion group ((1) of FIG. \ref{final_opinion}).

In contrast, with assessment rule 4, 7, 10 (KN), and 13, only players who share the same opinion on the recipient share the same opinion on the donor.
Because of this process, players who shared the same opinion on some players can become to share the same opinion on other players (e.g. when a player on whom they shared the same opinion play a lot of rounds in series as the recipient).
Finally, members who share the same opinion become same on all players.
In other words, players finally split into two opinion groups and players in different groups have the opposite opinion on all players.
Note, the members of each of the opinion groups are the members who have the same opinion on a player at the initial round.
At the initial state, every player is regarded as good by $pN$ players and as bad by $(1-p)N$ players on average.
Therefore, the size ratio of the two opinion groups is always $p:1-p$.
Then, opinion matrices can be sorted as (2) of FIG. \ref{final_opinion}.

Now we know whether all players belong to the same opinion group or they are split into two opinion groups with size ration $p:1-p$, for each assessment rule.
We can also determine the fraction of players who are regarded as good by themselves for each opinion group, at least if we fix the action rule.
Consider the case the assessment rule is KN and the action rule is DISC.
As mentioned above, assessment rule KN leads players to split into two opinion groups.
A player cooperate with who is good in his opinion and defect against who is bad in his opinion because of the action rule DISC.
In both cases, he and members of his opinion group will regard him as good because of the assessment rule KN. 
Therefore, the number of players who are good in their opinion monotonically increase.
Finally, all players are good in their own opinion in both opinion groups.
Similar procedures determine the stationary state of the opinion matrix for all combinations of an action rule and an assessment rule.

The stationary states of action matrices are clear from the action rule and the stationary state of opinion matrices.
In the example of KN and DISC, players in each opinion group regard only those in the same group are good, therefore they cooperate with only the members of their opinion group.
It directly means that they form two exclusive groups, with size ratio $p:1-p$, which is state (f).
Other cases go similarly.

Until here, we assumed all players adopt a common action rule and revealed that players split into two exclusive groups when the assessment is KN and the action rule is DISC or pDISC.
We finally study whether the state all players adopt DISC and the state all adopt pDISC are stable or not under the evolution of action rules, when the assessment rule is KN.
We consider the case that most of players adopt DISC (pDISC) and a slight fraction of players, ``mutants'', adopt another action rule and study whether the DISC players (pDISC players) get higher average payoff than the mutants.
If DISC players (pDISC players) get higher average payoff, DISC (pDISC) is called evolutionary stable strategy~\cite{Smith-Price} and it suggests the mutants' action rule do not spread.
Remember that $c$ is the cost which a donor takes to cooperate and $b$ is the benefit the recipient gains then.
We proved that when the assessment rule is KN, if and only if the benefit-cost ratio $b/c$ exceeds the threshold
\begin{eqnarray}
    \left(\frac{b}{c}\right)^{*}=\frac{1}{2}\left[\frac{1}{\left(2p-1\right)^{2}}+1\right],\label{eq:threshold}
\end{eqnarray}
DISC and pDISC are evolutionary stable strategies.
The threshold rapidly increases as initial trust probability $p$ approach $1/2$.
We will roughly see group separation is also accountable for this behavior of the threshold.

Consider the case almost all players adopt DISC and a few mutants adopt All-D.
All-D players hardly affect the relation among DISC players, therefore DISC players split into two opinion groups with size ratio $p:1-p$ as they do when DISC is the common action rule.
DISC players are cooperative inside of each group and not cooperative across the groups.
When the sizes of the two groups are more different, the chances for \emph{total} DISC players to be matched inside of the groups decrease, which means fewer chance of cooperation.
At the same time, All-D players gain higher payoff when DISC player split evenly. 
Owing to the disagreement between the two groups, All-D players always have chance to receive cooperation from either opinion groups.
If an All-D player defect against a player in an opinion group, the opinion group regards the All-D player as bad, however the other opinion group regards him as good.
If two opinion groups' sizes have a very large difference, All-D players almost always defect against members of majority group, therefore most of All-D players can receive cooperation only form the small minority group.
This is why DISC players' payoff can exceed All-D players' even with lower $b/c$, if $p$ is further from $1/2$.
The case of pDISC goes similarly.

Our results show that in-group favoritism occurs based on the spontaneously formed group structure, even without exogenous group structure.
Moreover, the separation into groups is caused by mutual assessment, which usually play a key role to overcome social dilemma.
Especially, it is surprising and interesting that KN, one of the best assessment rules to promote cooperation under public assessment, causes group separation under private assessment. 
This connection between the people's definitions of goodness and their group structure might be experimentally and empirically testable.

Extension of our model may enable us to understand not only completely separated two groups but also more complex and various group dynamics in our society.
For example, if we consider more complex assessment rules, with which observers take their original opinion on the donor into consideration, more than tow exclusive groups can be formed.
 
Formation of opinion groups plays the crucial role in the derivation of stationary state.
This analysis based on opinion groups can be applied to the system with small error and the case assessment rules are different among players, which enables, for example, analysis of evolution of assessment rules.

In conclusion, it was revealed that players split into two exclusive groups by specific types of reciprocity norm and that the norms can be evolutionary stable. 
This can be an explanation of why people frequently form groups and cooperate only inside of the groups.

\end{document}